\documentstyle[agupp]{article}

\twocolumn

\lefthead{Ruffolo and Khumlumlert}
\righthead{Coherent Pulses of Cosmic Rays}
\received{April 6, 1995}
\revised{June 2, 1995}
\accepted{June 20, 1995}
\journalid{GRL}{??? 1995}
\articleid{??}{??}
\paperid{95L00002N}
\ccc{???-???/95/95L00002N\$05.00}
\cpright{AGU}{1995}

\authoraddr{T. Khumlumlert and D. Ruffolo, Department of Physics,
Faculty of Science, Chulalongkorn University, Bangkok 10330,
Thailand.}

\slugcomment{Accepted by {\it Geophysical Research Letters}, 1995.}

\begin{document}

\title{Formation, Propagation, and Decay of Coherent Pulses of Solar
Cosmic Rays}

\author{D. Ruffolo and T. Khumlumlert}
\affil{Chulalongkorn University, Thailand}

\begin{abstract}
We have performed numerical simulations of the
interplanetary transport of solar cosmic rays.  The
particles form a coherent pulse within $\sim0.01$ AU
after their injection.  The gradual decrease of a
pulse's speed and anisotropy can be understood in terms of
an equilibrium between pitch-angle scattering and
focusing.  The results should be useful for estimating
times of particle injection.
\end{abstract}



\section{Introduction}

Much of the current research on solar cosmic rays aims to determine
the time distribution of their emission during solar flares.  The major
obstacle is that no spacecraft has approached closer than $\approx
0.3$ AU from the Sun, so interplanetary scattering has greatly broadened the
observed temporal distributions.  This problem is least severe when
the detector is magnetically reasonably well-connected to the flare.
If strong anisotropies are observed, they also
aid the determination of the scattering mean free path and the
injection profile (Kallenrode, Wibberenz, and Hucke 1992).

The key phenomenon underlying the strong aniso\-tropy observed for
many well-connected events is that charged particles of a given
energy tend to travel together as a highly anisotropic ``coherent
pulse,'' which can survive long enough to be observed near the Earth.
Figure 1 shows stages in the formation of such a pulse very close to the Sun.
This phenomenon was first systematically described by Earl (1974, 1976a, b)
who used analytic approximations to describe their formation and decay.
However, that description was limited by the assumption of a
constant focusing length, implying an
exponentially decaying magnetic field strength.
Here we re-examine the evolution of coherent pulses of solar cosmic
rays using numerical simulations for a more realistic Archimedean
spiral field (Parker 1958).

\section{Theory and Numerical Method}

For ions of $\gtrsim20$ MeV/n or electrons of $\gtrsim10$ keV, the transport
along the interplanetary magnetic field is well described by the
following equation (using the notation of Ng \& Wong 1979):
\begin{eqnarray}
\frac{\partial F(t,\mu,z)}{\partial t}&=&
  -\mu v\frac{\partial F(t,\mu,z)}{\partial z} \nonumber\\
& &\mbox{}-\frac{v}{2L(z)}\frac{\partial}{\partial\mu}[(1-\mu^2)F(t,\mu,z)]
  \nonumber \\
& &\mbox{} + \frac{\partial}{\partial\mu}\left[\frac{\varphi(\mu)}{2}
  \frac{\partial F(t,\mu,z)}{\partial\mu}\right],
  \label{eq:tr}
\end{eqnarray}
where
$z$ is the arclength along the magnetic field,
$\mu$ is the cosine of the pitch angle ($v_z/v$),
$F$ is the distribution function, defined as the
number of particles per $z$ per $\mu$ in a given magnetic flux tube,
$L$ is the focusing length, $B/(dB/dz)$, and
$\varphi$ is the coefficient of pitch-angle scattering.
The first term on the right hand side of eq.~(\ref{eq:tr})
expresses the streaming of particles along the magnetic field.
The second term is due to
pitch-angle scattering from irregularities in the magnetic field,
and the final term is due to adiabatic focusing (Roelof 1969).

We derive the focusing
length, $L(z)$, from an Archi\-med\-ean spiral field that makes a
45\deg\ angle with the radial direction at a radius of 1 AU,
corresponding to a solar wind speed of $\approx 400$ km/s.
The pitch-angle scattering coefficient, $\varphi(\mu)$, is
parameterized as
\begin{equation}
\varphi (\mu) = A|\mu|^{q-1}(1-\mu^2).
\end{equation}
following Jokipii (1971).  We express
our results in terms of the scattering mean free path, $\lambda$,
which is given by
\begin{equation}
\lambda=\frac{3}{(2-q)(4-q)}\frac{v}{A}.
\end{equation}
Note that for a constant value of $\lambda$, eq.~(\ref{eq:tr}) involves
only the distance traveled, $s=vt$, and not $v$ or $t$ alone.  We
will therefore show results in terms of $s$ instead of time.

The numerical method used to solve eq.~(\ref{eq:tr}) was essentially
the finite difference method of Ruffolo (1991), with some minor
improvements.  That study described tests of the code
and successful fits to the observed intensity and pitch-angle
distribution as a function of time for neutron-decay protons.
More recently, this and two completely different numerical
methods were shown to give very similar results (Earl et al.\ 1995).

\section{Results}

Numerical simulations were performed for an initial condition
corresponding to an instantaneous injection of particles near the Sun at $t=0$.
For most of the simulations, we took the initial pitch-angle distribution
to be concentrated at the highest $\mu$-grid value, and used an absorbing
boundary condition at $z=0$, i.e., $F(t>0,\mu>0,z=0)$ was set to zero.
(The injection site and inner boundary, $z=0$, was set to $r=0.01$ AU
$\approx2R_\odot$ to avoid a divergence in the focusing term.)
The outer boundary was set to the maximum value of $s=vt$.
Finally, the grid spacings
were $\Delta s=0.02$ AU, $\Delta\mu=0.08$, and $\Delta z=\Delta\mu
\Delta s=0.0016$ AU except where noted.

To examine the propagation of coherent pulses, we plot the
average distance along the magnetic field
of the particle distribution, $\langle z\rangle$, as a
function of $s=vt$ for selected values of $\lambda$ and $q$ (Figure~2).
The dashed line, $\langle z\rangle=s$, shows where the particles
would be if they all traveled directly along the magnetic field, with
$\mu=1$ and $v_z=v$.  As expected, the progress
along the magnetic field is fastest for the longest
mean free path, corresponding to the weakest scattering.

The rate of change of $\langle z\rangle$ with respect to $s$ is
\begin{equation}
\frac{d\langle z\rangle}{ds} = \langle\mu\rangle = \frac{\delta}{3},
\end{equation}
where $\delta$ is the anisotropy averaged over the particle
distribution, so $\langle\mu\rangle$ is proportional
to the pulse's propagation speed and anisotropy.
Figure~3a shows the evolution of $\langle\mu\rangle$
as a function of the average position, $\langle z\rangle$.
The most noticeable trend is that $\langle\mu\rangle$ monotonically
decreases with position in all cases, as the pulse moves to regions
where the focusing is weaker.  This slowing in turn causes the
curvature of the lines in Figure~2 away from the line of maximum speed.

Another noticeable feature in Figure~3a is the convergence of curves
for different $q$ toward a universal curve for smaller $\langle\mu\rangle$
values.  This convergence occurs later for higher values of
$\lambda$ and always appears at similar values of $\langle\mu\rangle$,
indicating that the convergence coincides with the
decay of the coherent pulse toward a diffusive distribution.
This can be verified by examining $\sigma_z=\sqrt{\langle(z-\langle
z\rangle)^2\rangle}$
(Figure~3b).  We note that the position at which
$\sigma_z=\langle z\rangle/4$ (long-dashed line) for each $\lambda$
corresponds roughly to the position of convergence, and to a value of
$\langle\mu\rangle\approx 1/3$.  We will therefore
refer to $\langle\mu\rangle\gtrsim 1/3$ as the coherent pulse r\'egime
and to $\langle\mu\rangle\lesssim 1/3$ as the diffusive r\'egime.

The behavior of $\langle\mu\rangle$ vs.\ $\langle z\rangle$
can be interpreted in terms of an
equilibrium between the scattering and focusing terms in eq.\
(\ref{eq:tr}) with the streaming term ``turned off.''
[The scattering/focusing eigenfunctions
defined by Earl (1976a) used a different form of eq.~(\ref{eq:tr}),
and thus have a different physical interpretation.]
Figure~4 shows simulation results for the equilibrium average of
the pitch-angle cosine, $\langle\mu\rangle_{eq}$, and the
equilibration distance, $s_{eq}$, which is the
distance traveled over which equilibrium is achieved, or $v$ times the
longest relaxation time.  The results for $\langle\mu\rangle_{eq}$
are in good agreement with the analytic expression
\begin{equation}
\langle\mu\rangle_{eq}=
\frac{\int_{-1}^1\mu F_{eq}d\mu}{\int_{-1}^1F_{eq}d\mu}=
\frac{\int_{-1}^1\mu \exp
  \left(\case{v}{AL}\case{\mu|\mu|^{1-q}}{2-q}\right)d\mu}
{\int_{-1}^1\exp
  \left(\case{v}{AL}\case{\mu|\mu|^{1-q}}{2-q}\right)d\mu}
\end{equation}
(fortunately, for the $q$ values chosen
here, these integrals can be solved in terms of simple functions).
For large $z$ (i.e., $L\gg\lambda$, or the weak focusing limit),
$s_{eq}$ approaches a $q$-dependent value (equal to or slightly
larger than $\lambda$) which agrees with the $L=\infty$ value of
Bieber (1978).

For very small values of $z$, where there is very strong focusing,
$s_{eq}$ is also small, and the distribution achieves pitch-angle
equilibrium very quickly.  At the same time,
$\langle\mu\rangle_{eq}\approx 1$, so the
distribution ``relaxes'' rapidly (within $\sim0.01$ AU)
to form a coherent pulse with
nearly maximal anisotropy and highly collimated motion along the
magnetic field (Figures 1 and 5; a finer grid spacing
was used for these simulations).  Note that the spiky features in
Figure 1, before pitch-angle equilibrium is reached, are due to the
low scattering coefficient, $\varphi(\mu)$, near $\mu=1$.
Initial injections that were
isotropic or highly focused yielded essentially the same distribution
beyond $z=0.03$ AU, with the former lagging by only 0.004 AU.  A
reflecting boundary condition gave virtually identical results.

Farther from the Sun, we see
that $\langle\mu\rangle_{eq}$ rapidly declines with $z$ (Figure 4).
Since the equilibration distance rapidly increases, the pitch-angle
equilibrium is not maintained and $\langle\mu\rangle$ declines less
rapidly.  For a narrow pulse, the deviation from pitch-angle equilibrium
evolves according to
\begin{equation}
\frac{d(\langle\mu\rangle-\langle\mu\rangle_{eq})}{ds}
  = -\frac{\langle\mu\rangle-\langle\mu\rangle_{eq}}{s_{eq}}
  -\langle\mu\rangle\frac{d\langle\mu\rangle_{eq}}{dz}.
\label{eq:dev}
\end{equation}
However, this equation gives a steady-state deviation which is
insufficient to account for the difference between $\langle\mu\rangle$
and $\langle\mu\rangle_{eq}$ in the diffusive r\'egime.  The
remaining deviation can be explained by considering the width of the
distribution.  During the diffusive phase, the average of
$\langle\mu\rangle_{eq}$ over the
distribution can deviate significantly from the value at $\langle z\rangle$.
Thus we see that the convergence of $\langle\mu\rangle$ as a function
of $\langle z\rangle$ for different $q$'s in the diffusive phase is due to
1) the convergence of $\langle\mu\rangle_{eq}(z)$ toward a
$q$-independent function for $\lambda\ll L$, which can be shown to be
$\case{1}{3}\lambda/L(z)$, and 2) the similarity of $\sigma_z$ as a function
of $\langle z\rangle$ for the three values of $q$ at each $\lambda$
(Figure~3b).

\section{Discussion}

We note that the transport equation used here neglects several
well-known effects.  Diffusion and drifts perpendicular to the field
could reduce the intensity, especially for well-connected events, for
which neighboring flux tubes have a lower intensity, but this should
not significantly affect the propagation diagnostics discussed here.
Adiabatic deceleration
and convection were included in numerical simulations by Ruffolo
(1995) and were shown to noticeably affect the propagation for ions
of $\lesssim20$ MeV/n and electrons of $\lesssim10$ keV.

On the other hand, this approach does overcome the main limitation
of the previous, analytic descriptions of coherent pulses (Earl
1976a, b), i.e., the assumption of a constant focusing length
($B\propto e^{-z/L}$).  We find that several of their conclusions do not
apply for the case of a realistic Archimedean spiral field.  We find that
for this case, the
propagation speed is not constant, and does not exhibit distinct
supercoherent and coherent r\'egimes, but rather decreases steadily
along with $\langle\mu\rangle_{eq}$.  Other results from our
simulations show that the pulse width
increases roughly linearly with time, in contrast with the
result for an exponential field that $\sigma_z\propto t^{1/2}$ (Earl 1974).
Of course, these limitations do not diminish the importance of that
seminal work.

It is hoped that these results will be applicable to studies
of the time profile of particle injection from solar flares,
or comparisons of the time of injection for different particle species.
When a strong coherent pulse is observed, one can estimate $\lambda$
from the anisotropy and the pulse width (Figure 3).  A lack of
consistency could indicate a finite injection width (if the pulse
width is high), or shock acceleration outside the corona (if the
anisotropy is high).  One can derive the
propagation time from $\lambda$, given an assumption or measurement for $q$.
Subtracting the propagation time from the time of maximum then yields
an estimated time of injection.  Note that conditions with $\lambda\sim 1$
AU, corresponding to the
highest mean free paths reported in the inner heliosphere,
are sometimes called ``scatter-free'' conditions, with the
implication that particles freely stream along the field
at their maximum speed.  However, our results show that
even then, scattering significantly delays the arrival of the
bulk of the pulse, while the ``onset'' is delayed somewhat less (see
also Kallenrode and Wibberenz, 1990).  In general, Figure 2 can be used to
derive an accurate estimate of the propagation time.

\acknowledgments
The authors would like to\newline
thank the Laboratory for Astrophysics and
Space Research at the University of Chicago for kindly allowing
remote access to their computing facilities.

\newpage



\begin{figure}
\figurewidth{18pc}
\caption{Three stages in the evolution of a coherent pulse:
the distribution function, $F$ (vertical direction), vs. distance along
the magnetic field, $z$, and pitch-angle cosine, $\mu$, for
$s=vt=0.002$, 0.01, and 0.02 AU, when $\lambda=0.3$ AU, $q=1.5$,
and $F(t=0)$ is concentrated at $z=0$ and uniform in $\mu$.
}
\end{figure}

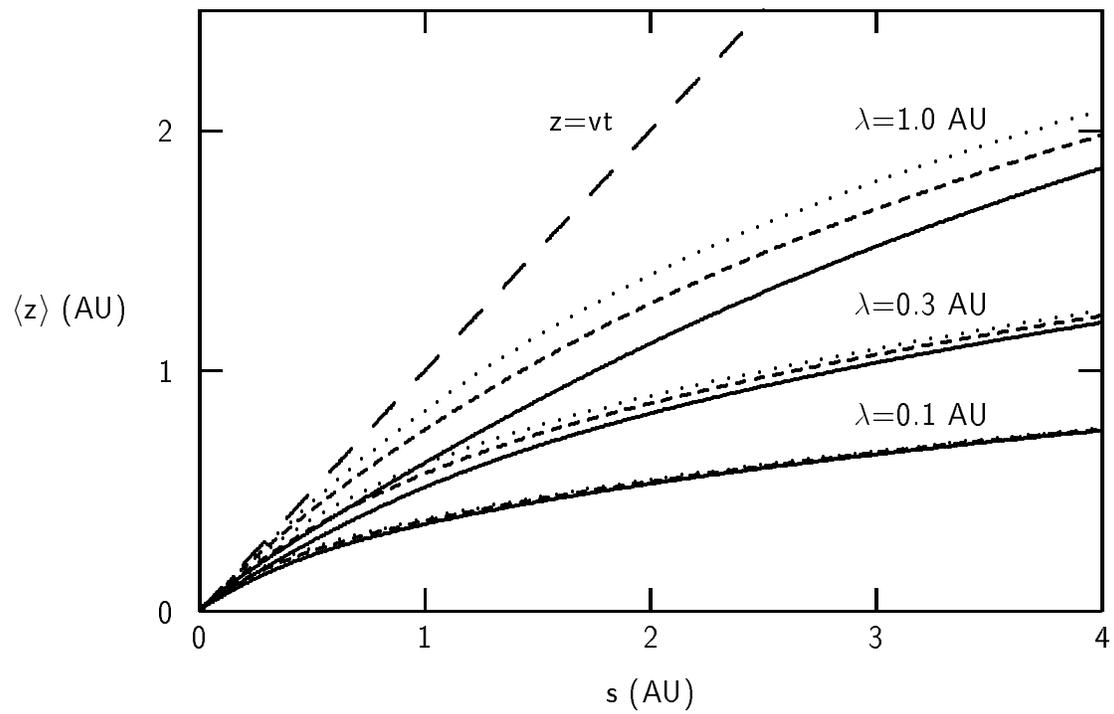
\begin{figure}
\figurewidth{18pc}
\caption{Simulation results for the
mean distance along the magnetic field, $\langle z\rangle$,
vs.\ the distance
traveled, $s=vt$, for $q=1.0$ (dotted lines), $q=1.5$ (dashed lines),
and $q=1.9$ (solid lines) and for the indicated values of $\lambda$.
The long-dashed line corresponds to motion directly along the
magnetic field.
}
\end{figure}

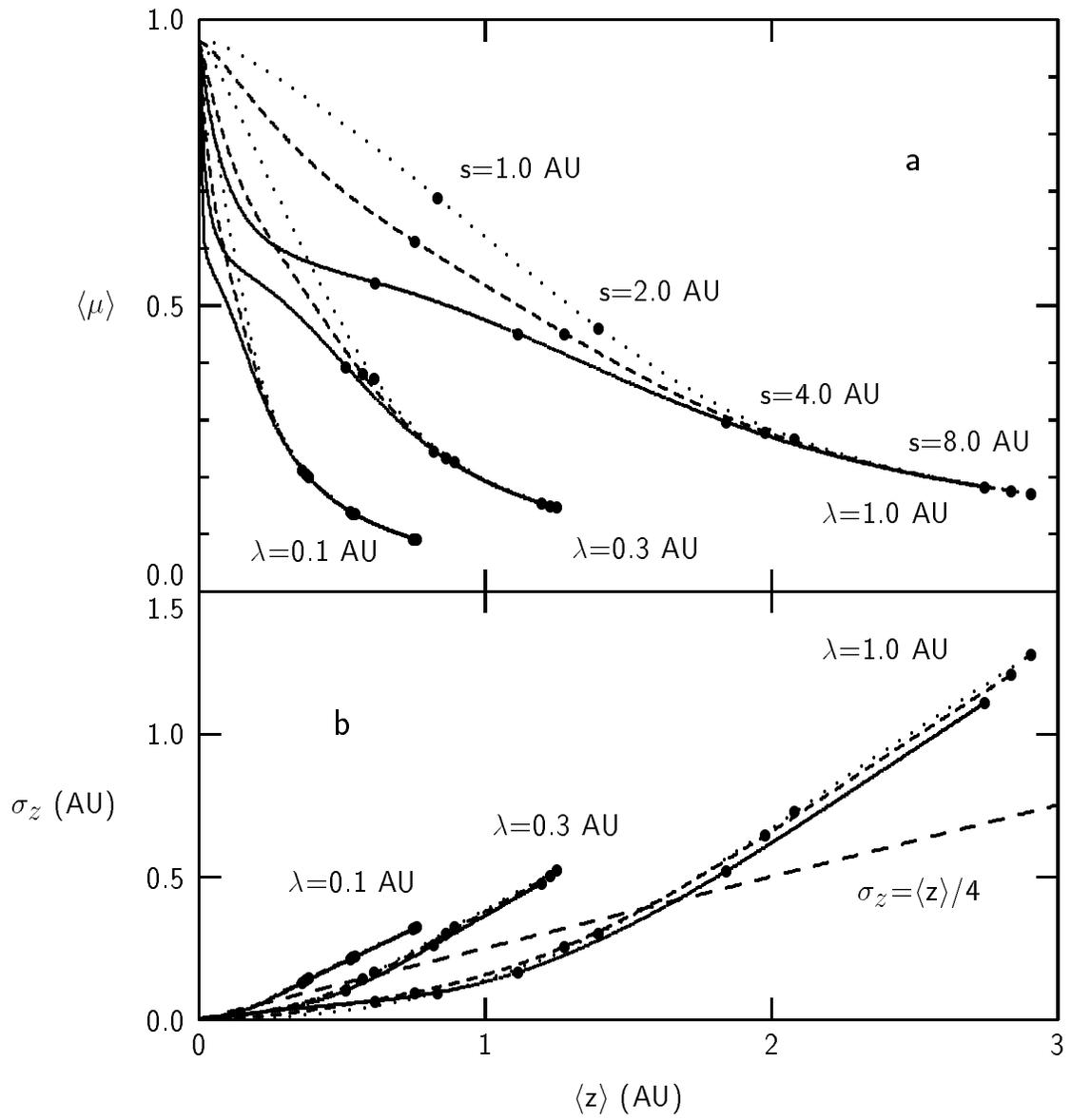
\begin{figure}
\figurewidth{18pc}
\caption{a) Mean pitch-angle cosine, $\langle\mu\rangle=\langle
v_z\rangle/v$, and b) pulse width, $\sigma_z$, vs.~mean
distance along the magnetic field, $\langle z\rangle$.
For an explanation of the curves, see Figure~1.  Solid circles
show points on each curve for specific values of $s=vt$.
Note that only curves for $\lambda=1.0$ AU
are extended to $s=8.0$ AU.
}
\end{figure}

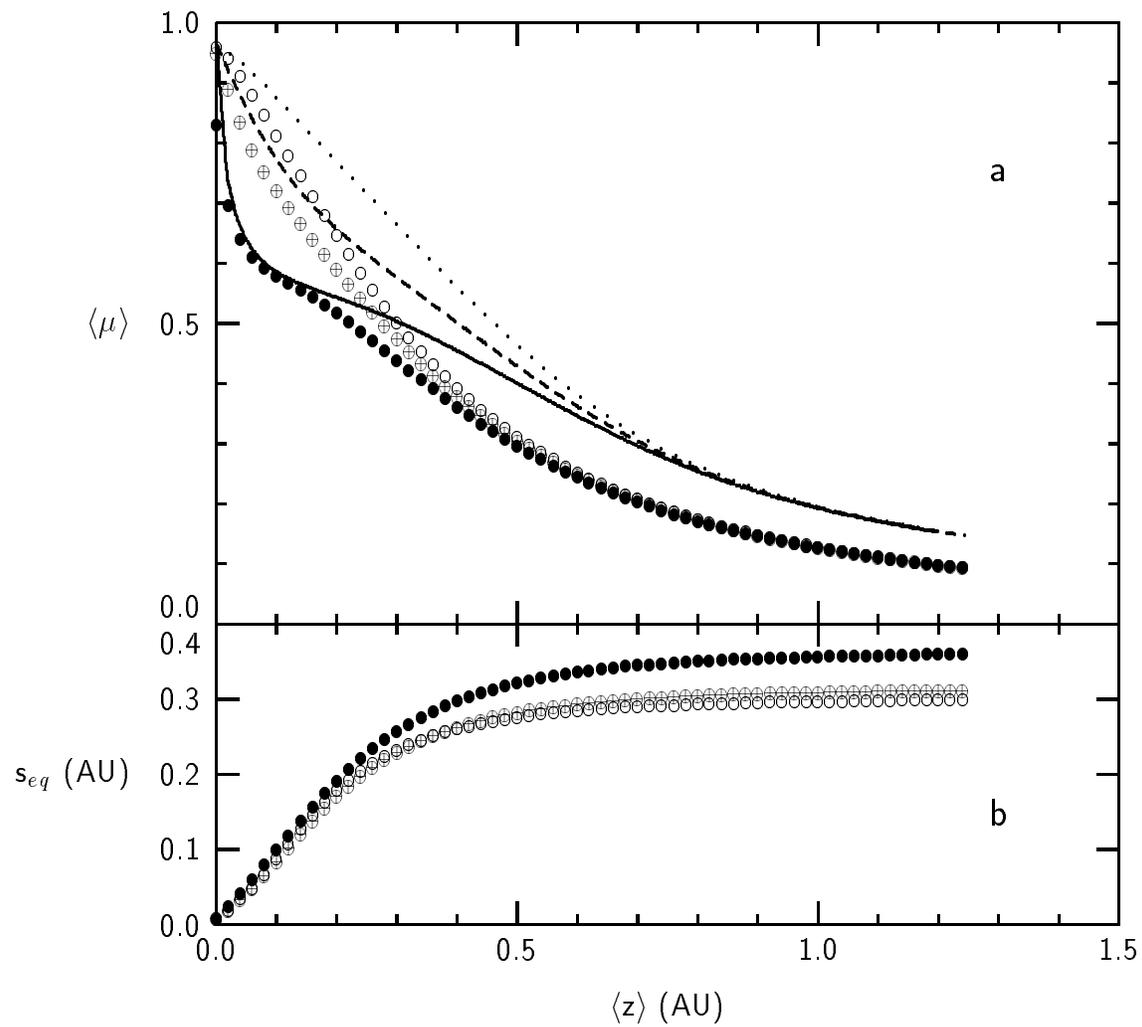
\begin{figure}
\figurewidth{18pc}
\caption{a) Mean pitch-angle cosine, $\langle\mu\rangle$, vs.~mean
distance, $\langle z\rangle$, up to $s=4$ AU for $\lambda=0.3$ AU
and $q$ values as specified for Figure~1, compared with
$\langle\mu\rangle_{eq}$ for $\lambda=0.3$ and $q=1.0$ ($\circ$), 1.5
(\lower-0.5ex\hbox{\fivsy\char'010}), and 1.9 ($\bullet$).  b) Equilibration
distance, $s_{eq}$, vs.~$\langle z\rangle$ for $\lambda$ and $q$ as
above.
}
\end{figure}

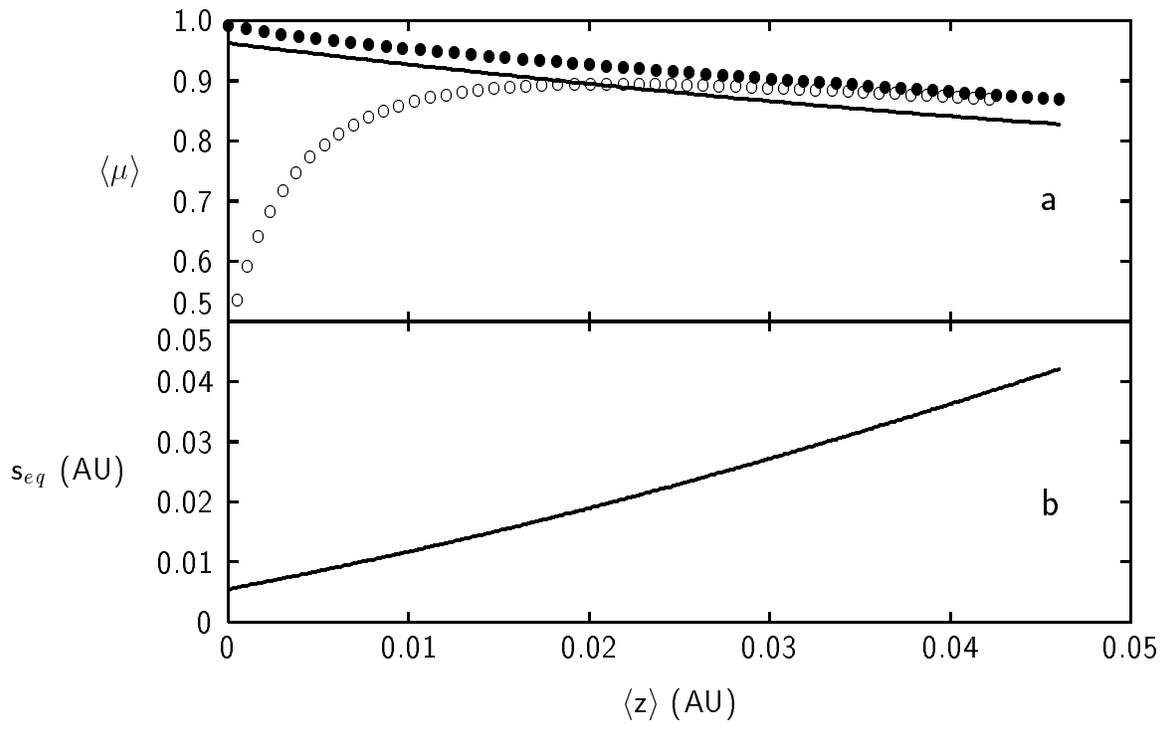
\begin{figure}
\figurewidth{18pc}
\caption{a) Mean pitch-angle cosine, $\langle\mu\rangle$, vs.~mean
distance, $\langle z\rangle$, up to $s=0.05$ AU for $\lambda=0.3$ AU,
$q=1.5$, and
isotropic ($\circ$) and highly focused ($\bullet$)
initial distributions, compared with $\langle\mu\rangle_{eq}$ (solid
curve). b) Equilibration distance, $s_{eq}$, vs.~$\langle
z\rangle$.
}
\end{figure}

\end{document}